\documentstyle[aps,preprint,psfig]{revtex}
\tighten

\newcommand{\beq}{\begin{equation}}
\newcommand{\eeq}{\end{equation}}
\newcommand{\beqa}{\begin{eqnarray}}
\newcommand{\eeqa}{\end{eqnarray}}
\newcommand{\etal}{{\it et al.}}

\newcommand\ph{\phantom{-}}
\def\nue{{\nu_e}}

\newcommand{\dm}{\mbox{$\Delta{m}^{2}$~}}

\newcommand{\br}{\mbox{$^{8}$B~}}
\newcommand{\ber}{\mbox{$^{7}$Be~}}

\def\js{Just So$^2$~}
\begin{document}


\title{Global oscillation analysis of solar neutrino data 
with helioseismically constrained fluxes}

\author{Sandhya Choubey$^1$\thanks{e-mail: sandhya@theory.saha.ernet.in},
Srubabati Goswami$^1$\thanks{e-mail: sruba@theory.saha.ernet.in},
Kamales Kar$^1$\thanks{e-mail: kamales@tnp.saha.ernet.in},
H.M. Antia$^2$\thanks{e-mail: antia@astro.tifr.res.in}, 
S.M. Chitre$^{2,3}$\thanks{e-mail: chitre@astro.tifr.res.in}\\
$^{1}${\it Saha Institute of Nuclear Physics,1/AF, Bidhannagar,
Kolkata 700 064, India}\\
$^2${\it Tata Institute of Fundamental Research,
Homi Bhabha Road, Mumbai 400005, India}\\
$^3${\it Queen Mary and Westfield College, Mile End Road, London E1 4NS, U. K.}}
\maketitle


\begin{abstract}
A seismic model for the Sun calculated using
the accurate helioseismic data
predicts a lower $^{8}{B}$ neutrino flux 
as compared to the standard solar model (SSM). However, there persists a
discrepancy between the predicted and measured neutrino fluxes and it seems
necessary to invoke neutrino oscillations to explain the measurements.
In this work, we have performed
a global, unified oscillation analysis of the latest solar
neutrino data (including the results of SNO charged current rate) 
using the seismic model fluxes as theoretical predictions.
We determine the best-fit values of the neutrino oscillation parameters 
and the $\chi^2_{\mathrm min}$ for both
$\nu_e-\nu_{\mathrm active}$ and $\nu_e -\nu_{\mathrm sterile}$ cases and
present the allowed parameter
regions in the $\Delta m^2 - \tan^2 \theta$ plane for
$\nu_e-\nu_{\mathrm active}$ transition.
The results are compared with those obtained using the latest
SSM by Bahcall and his collaborators.
  
\end{abstract}


\section{Introduction} 

Solar neutrino fluxes measured by all the experiments to date 
are
significantly at variance with the expected theoretical predictions. 
The most recent confirmation of this has come from  the heavy water (D$_{2}$O)
detector at 
Sudbury Neutrino Observatory (SNO) \cite{sno} 
which measures the 
solar \br neutrinos through the charged current (CC) 
reaction $\nu_e d\rightarrow p p e^{-}$. 
SNO has also published their result of \br flux measured by the 
neutrino-electron scattering reaction (ES) and reported a lower 
$^{8}$B flux as compared to theoretical 
predictions of the standard solar model. 
This is in agreement with the \br flux measured by the 
SuperKamiokande (SK) detector through the same reaction \cite{sksolar}. 
Thus, SNO and SK confirm the deficit of solar neutrino fluxes reported first in 
the $^{37}$Cl
radiochemical experiment of Davis {\it et al.} \cite{cl} and  
subsequently by  
Kamiokande \cite{kam} and the 
radiochemical $^{71}$Ga experiments
SAGE, GALLEX and GNO \cite{ga}. The theoretical predictions most widely used 
are from the standard solar model (SSM)
developed and remarkably refined over the last four decades
by Bahcall and his collaborators \cite{bah,bp98,bp00}.  
In recent years, the observations of solar oscillations have provided an
independent test of solar models. Inversions of accurately
measured frequencies of
solar oscillations have enabled a determination of the sound speed and density
profiles inside the Sun\cite{dog96}. While the SSM matches these inverted
profiles remarkably well, there is still a significant discrepancy
which is much larger than the errors in helioseismic inversions.
It would, therefore, be desirable to check the results on neutrino oscillation
solutions to the solar neutrino problem using a solar model which is
consistent with the helioseismic data. Such models can be constructed
using the inverted profiles of sound speed and density along with the
equations of thermal equilibrium, provided the heavy element abundance
profile and the input physics like opacity, equation of state and
nuclear reaction rates are assumed to be known \cite{shi96,ant98}.
These models represent the present Sun and do not depend on evolutionary
history of the Sun.
Such seismic models can be used to calculate solar neutrino fluxes
which turn out to be somewhat different from those obtained
with the standard solar model. 

In this work,
we consider the seismic model calculated using the technique
described by Antia and Chitre \cite{ant98}, but using updated helioseismic
data. This model
predicts a lower $^{8}$B flux than the BP2000 SSM \cite{bp00}.
However, when all the experimental rates are taken together there is still
inconsistency between theory and experiment \cite{ant98}.
This inconsistency cannot be removed even if opacities and heavy element
abundances are varied by arbitrarily large amount \cite{ant97}.
Thus one needs to invoke neutrino oscillations to explain the observed
fluxes of solar neutrinos. Seismic models have also provided some
constraints on the pp reaction cross-section \cite{ant98,inn98,ant99,sch99}.
It appears that pp reaction cross-section needs to be increased by about
4\% over the value used by Bahcall et al. \cite{bp00}, to obtain solar models
that are consistent with seismic data. This increase in pp reaction rate
in the seismic model with correct luminosity 
tends to decrease the predicted neutrino fluxes 
for all four experiments.

In addition to the data on the total flux, SK also provides the data 
on the day-night recoil electron energy spectrum \cite{sksolarspec}
and the zenith angle distribution of events \cite{sksolardn}.
SNO has also reported the recoil electron energy spectrum for the 
CC events. 
Global oscillation analysis of the data on rates and spectrum has
been carried out by different groups to put constraints on the 
oscillation parameters \cite{bks98,valle,gmr1,gmr2,bks2001,dp} (pre-SNO)
and \cite{flsno,bcc,ab,strumia} (post-SNO)
These studies  have used the neutrino
fluxes from the standard solar model of Bahcall {\it et. al.}~\cite{bp98,bp00}.  
However, it would be interesting to see the implications of  
neutrino fluxes from the seismic model on the oscillation parameters.
In this work, we perform a global and unified oscillation analysis of the
solar neutrino data on total fluxes from the SNO, Cl, Ga and SuperKamiokande 
experiments 
together with the day-night spectrum data from SK using the seismically 
inferred  neutrino fluxes 
and compare the results with 
those obtained using the latest standard solar model (BP2000) of 
Bahcall {\it et. al.} \cite{bp00} \footnote{We have not incorporated the 
SNO ES data as well as the SNO CC spectrum data as they still have large 
errors.}. 
We use the latest 1258-day SK data in our analysis \cite{sk1258}. 

The rest of the paper is organized as follows:
in section 2 we present the basic features of the 
seismic model and describe the  
formalism for the unified analysis in Section 3.
In section 4 we discuss the procedure for the analysis and the results 
and finally, summarize the conclusions in Section 5.

\section{The Seismic Model} 
We use the mean frequencies from the Michelson Doppler Imager (MDI)
data from the first 360 days of its operation \cite{mdi97}, to calculate the sound
speed and density profiles inside the Sun using a regularized least squares
inversion technique \cite{ant96}.
We adopt the inverted sound speed and density
profiles, along with the heavy element abundance ($Z$) profile
from the solar model of \cite{rich96} to calculate the
seismic model using equations of thermal equilibrium
\cite{ant98}. We use the OPAL opacities 
\cite{igl96} and equation of state \cite{rog96},
and nuclear reaction rates from \cite{adel98},
except for the proton-proton reaction rate, which is slightly
adjusted to
yield the correct observed solar luminosity \cite{ant99}.
With the latest physical and seismic input, the pp reaction cross-section
$S_{11}$ needs to be increased to $4.17\times10^{-25}$ MeV barns.
The helium abundance by mass at the base of the convection zone
in this seismic model turns out to be 0.250, which is in good agreement
with the helioseismic estimate in the envelope \cite{basu98}.

With the knowledge of temperature, density and composition
profiles from the seismic model, neutrino fluxes can be calculated
and the results are shown in Table \ref{tab1}. Apart from inversion errors,
other main sources of uncertainties in these calculated fluxes are
the nuclear reaction cross-sections for the $^3$He-$^3$He ($S_{3,3}$),
$^3$He-$^4$He ($S_{3,4}$), p-$^{14}$N ($S_{1,14}$), p-$^7$Be
($S_{1,7}$) reactions, as well as the solar luminosity, the heavy
element abundance, $Z$
and opacities, $\kappa$. To estimate the effect of these quantities
on neutrino fluxes in the seismic model we calculate the logarithmic
derivatives of neutrino fluxes with respect to each of these
quantities ($X_i$) and these are also listed in Table \ref{tab1}, with
the last row showing the estimated
relative errors in these quantities.
Apart from these we also include the uncertainty
due to the electron capture rate of the 
process  
$^{7}$Be($e^{-},\nu_e)^{7}$Li and the astrophysical uncertainty in the 
$S_{0}$ factor of the reaction $^{16}$O($p,\gamma)^{17}$F.  
These contributions are same as in \cite{bp00}. 
The expected neutrino fluxes in various solar
neutrino experiments can be calculated from Table \ref{tab1} and these
values are given in Table \ref{tab2}, for comparison with observed
fluxes \cite{cl,ga,sk1258} and those in the 
standard solar models \cite{bp00,bp98,bru98,bru99}.
In Table \ref{tab3} we present the 
contributions of the various neutrino sources 
to the Cl and Ga experiments according to the seismic model and BP2000.  
Table \ref{tab4} summarises the ratios of the experimental rates to the 
theoretical predictions for the Cl, Ga, SK and SNO experiments 
for both BP2000 and seismic model. For SNO we display only the 
CC rate. 
We also show the compositions of major flux components. 

The neutrino fluxes in our seismic model are somewhat different from
those in the seismic model of Watanabe and Shibahashi \cite{seismic}.
The main difference arises because they have used only the sound speed
from primary inversions to calculate the seismic model, while we have
used both sound speed and density profiles from primary inversions.
Thus the
density profile in seismic model of Watanabe and Shibahashi does not,
in fact, match
the inverted density profile. Further, since the density profile in their
model is not constrained to seismic profile they are able to construct
seismic model with correct luminosity using the standard pp nuclear
reaction rate, which tends to give larger neutrino fluxes. If they
were to assume a slightly larger cross-section for pp reaction, the
density profile in their model would be in better agreement with the
inverted profile. Apart from this, the adopted
heavy element abundance, $Z$-profile is also different in their
work. While we have taken the $Z$-profile
from model of Richard et al. \cite{rich96}, which includes mixing just below
the base of convection zone, as implied by helioseismic data
\cite{basu94,dog96}, Watanabe and Shibahashi \cite{seismic}
use a homogeneous $Z$ profile.
Similarly, the neutrino fluxes in our seismic model
are somewhat smaller than those in SSM \cite{bp00}. The main reasons for
this reduction are again the increase in $S_{11}$ and neglect of
mixing just beneath the convection zone in SSM. If these had been
incorporated, then the SSM fluxes should also come close to
the corresponding seismic model fluxes.

\section{Formalism for Unified Oscillation Analysis}
The general expression for the probability amplitude of survival for an
electron neutrino produced in the deep interior of the Sun, for
two neutrino flavors, is given by \cite{qvo}
\begin{equation}
A_{ee} = A_{e1}^\odot  A_{11}^{\mathrm vac} 
A_{1e}^\oplus + A_{e2}^\odot A_{22}^{\mathrm vac}
A_{2e}^\oplus
\label{amp}
\end{equation}
where $A_{ek}^\odot (k=1,2)$ gives the probability amplitude of
$\nu_e \rightarrow \nu_k$
transition at the solar surface,
$A_{kk}^{\mathrm vac}$ is the survival amplitude from the solar surface to the
surface of the Earth and
$A_{ke}^\oplus$ denotes the $\nu_k\rightarrow \nue$
transition amplitudes inside the Earth.
We can express 
\begin{equation}
A_{ek}^\odot = a_{ek}^\odot e^{-i \phi^\odot_{k}}
\eeq
where $\phi^{\odot}_k$ is the phase picked up by the neutrinos
on their way from the production point in the central regions to the
surface of the Sun and
\beq
{a_{e1}^{\odot}}^2 = \frac{1}{2} + (\frac{1}{2} - P_J)\cos2\theta_m
\label{ae1}
\eeq
$\theta_m$ being the mixing angle at the production point
of the neutrino, given by 
\begin{equation}
\tan 2\theta_{m} =  \frac{\Delta m^2\sin 2\theta}{\Delta m^2\cos2\theta -
2\sqrt{2}G_{F}n_{e}E}.
\label{thetam}
\end{equation}
Here $n_{e}$ is the ambient electron density, $E$ the
neutrino energy, and \dm (= ${m_{2}^2 - m_{1}^2}$) is the mass squared
difference in vacuum. We denote by
$P_{J}$ the non-adiabatic level jumping probability
between the two mass eigenstates which for an exponential density 
profile\footnote{Note that for the actual calculation of the probabilities 
we have used the numerical density 
profile given in \cite{bp00} for SSM and the model in \cite{ant98} with updated helioseismic data for seismic model.} can be expressed as 
\cite{petcov}
\beqa
P_J = \frac{\exp(-\gamma \sin^2 \theta) - \exp(-\gamma)}{1-\exp(-\gamma)}
\eeqa
\beqa
\gamma =\pi\frac{\Delta m^2}{E}\left| \frac{d~\ln n_e}{dr}
\right |_{r=r_{res}}^{-1}
\eeqa
For sterile neutrinos, the $n_e$ in Eq. (\ref{thetam}) has to be 
replaced by $n_e - \frac{1}{2}n_n$, where $n_n$ is the neutron number 
density. The survival amplitude 
$A_{kk}^{\mathrm vac}$ is given by
\beq
A_{kk}^{\mathrm vac} = e^{-i E_{k} (L - R_{\odot})}
\eeq
where $E_k$ is the energy of the state $\nu_k$, $L$ is the distance
between the center of the Sun and Earth and $R_\odot$ is the solar
radius.
For a two-slab model of the Earth --- a mantle and core with constant
densities of 4.5 and 11.5 gm cm$^{-3}$
respectively, the expression for $A_{2e}^{\oplus}$ can be written as
(assuming the flavor states to be continuous across the boundaries)
\cite{petearth}
\beqa
A_{2e}^{\oplus}&=&
\sum_{\stackrel{i,j,k,}{\alpha,\beta,\sigma}}
U_{ek}^M e^{-i\psi_k^M}
U_{\alpha k}^M U_{\alpha i}^C
e^{-i\psi_i^C}
U_{\beta i}^C U_{\beta j}^M
e^{-i\psi_j^M}
U_{\sigma j}^M U_{\sigma 2}
\eeqa
where ($i,j,k$) denotes mass eigenstates and 
($\alpha,\beta,\sigma$) denotes flavor eigenstates, $U$, 
$U^M$ and $U^C$ are the mixing matrices in vacuum, in the mantle and the
core respectively and $\psi^M$ and $\psi^C$ are the corresponding phases
picked up by the neutrinos as they travel through the mantle and the core
of the Earth.
The $\nu_e$ survival probability is given by 
\begin{eqnarray}
P_{ee}  & = &  |A_{ee}|^2 \nonumber \\
        & = & {a_{e1}^\odot}^2 |A_{1e}^\oplus|^2
+ {a_{e2}^\odot}^2 |A_{2e}^\oplus|^2
                \nonumber \\
        &   & + 2 a_{e1}^\odot a_{e2}^\odot
               Re[A_{1e}^\oplus {A_{2e}^\oplus}^{*}
e^{i(E_{2} - E_{1})(L - R_{\odot})} e^{i(\phi_{2,\odot} - \phi_{1,\odot})}]
\label{pr}
\end{eqnarray}
Identifying 
$P_{\odot} = {a_{e1}^\odot}^2$ and 
$P_{\oplus} = |A_{1e}^\oplus|^2$
Eq. (\ref{pr}) can be expressed as \cite{stp,murayama,qvo}
\beqa
P_{ee}&=&P_{\odot}P_{\oplus} + (1-P_{\odot})
(1-P_{\oplus})\\
&& + 2\sqrt{P_{\odot}(1-P_{\odot})
P_{\oplus}(1-P_{\oplus})}\cos\xi
\label{probtot}
\eeqa
where we have combined all the phases involved in the Sun, vacuum and
inside Earth in $\xi$.
This is the most general expression for survival probability for the 
unified analysis of solar neutrino data. 
Depending on the value of $\Delta m^2/E$ one recovers the well known 
{\it Mikheyev-Smirnov-Wolfenstein} (MSW) \cite{msw}
and vacuum oscillation (VO) \cite{vac} limits: 
\begin{itemize} 
\item In the regime $\Delta m^2/E \stackrel{<}{\sim}
5\times 10^{-10}$ eV$^2$/MeV matter effects inside the Sun  
suppress flavor transitions and 
$\theta_m \approx \pi/2$. Therefore, from (\ref{ae1}), we obtain 
$P_{\odot} \approx P_J \approx \cos^2 \theta$
as the propagation of neutrinos is extremely 
non-adiabatic and likewise,
$P_{\oplus} = \cos^2\theta$ to give 
\beq
P_{ee}^{\mathrm vac} = 1 - \sin^2 2\theta \sin^2(\dm (L-R_\odot)/4E)
\label{probvac}
\eeq

\item For $\Delta m^2/E \stackrel{>}{\sim} 10^{-8}$
eV$^2$/MeV, the total oscillation phase 
becomes very large and the $\cos\xi$ term in Eq. (\ref{probtot})
averages out to zero. 
One then recovers the usual 
MSW
survival probability
\beq
P_{ee}^{\mathrm MSW} = P^{D} + 
\frac{(2P^D
-1)(\sin^2\theta- |A_{2e}^{\oplus}|^2)}
{\cos2\theta}
\label{probnight}
\eeq
The day-time probability is given by $P^{D}$ being 
\beq
P^{D} = \frac{1}{2} + (\frac{1}{2} - P_J) \cos2\theta\cos2\theta_m
\label{probday}
\eeq

\item In between the {\it pure} vacuum oscillation regime where the
matter effects can be safely neglected, and the {\it pure} MSW
zone where the coherence effects due to the phase $\xi$ can
be conveniently disregarded, is a region where both effects can
contribute. For
$5\times 10^{-10}$ eV$^2$/MeV $\stackrel{<}{\sim}
\Delta m^2/E \stackrel{>}{\sim} 10^{-8}$ eV$^2$/MeV, both matter effects
inside the Sun and coherent oscillation effects in the
vacuum become important. This is the {\it quasi vacuum oscillation}
(QVO) regime \cite{qvo}. In this region, $P_\odot\approx P_J$ and 
$P_\oplus=\cos^2\theta$ and the survival probability is given by 
\cite{stp,flqvo}
\beq
P_{ee} = P_J\cos^2\theta + (1-P_J)\sin^2\theta 
+\sin^22\theta \sqrt{P_J(1-P_J)}\cos\xi
\eeq
we calculate $P_J$ in this region using the 
prescription given in \cite{flqvo}.
\end{itemize}

\section{Results and Discussions}

In this section we present our results of global 
$\chi^2$ -analysis of the data on total rates and 
the day-night spectrum observed in SK.  
We have done two sets of calculations taking the theoretical 
predictions from the seismic model as well as  from BP2000 SSM. 
Our principal objective is to compare the two sets of results. 

Different approaches have been adopted regarding the treatment of the 
SK data on total rates and the recoil 
electron energy spectrum, in the global analysis. 
A critical  comparison of the various methods used and the dependence of 
the final results on the method followed is discussed in a lucid and 
extensive manner in \cite{bks2001}. Below we summarise the salient features.   
\begin{itemize}
\item (i) One  approach is to  
to include both the SK rate and the spectrum in the global analysis 
\cite{gmr2,bks2001}. 
\item(ii) The total rate measured in SK is not independent 
of the spectrum and hence the above method of including both
may lead to a possible overcounting of events. 
To avoid this the total SK rate is excluded from the global analysis 
in \cite{bks2001,bcc}. 
\item (iii) Another approach adopted to avoid the overcounting is to  
include the total SK rate in the global analysis 
but to adopt a free normalization factor for the spectrum 
\cite{valle,bks2001,dp,flsno,ab}.
Thus the spectrum gives information on only  
the shape of the \br survival probability. 
\end{itemize}
In case of methods (i) and (ii),
if the \br flux normalisation is varied as a free parameter in both rates 
and spectrum, no error correlation between these needs to be 
taken \cite{bks2001}. 
On the other hand, if the \br flux normalisation is kept fixed at 
the SSM value, then one should incorporate the correlation between 
the error in the 
the \br flux measured in the total rates and the SK spectrum \cite{gmr2,bcc}. 
For method (iii), since the normalisation of the spectrum is varied as 
a free parameter, the $\chi^2$ due to rates and spectrum can be summed 
independently.  
For the purpose of this paper we adopt method (iii), so that our 
total $\chi^2$ is given by
\beq
\chi^2_{\mathrm total} = \chi^2_{\mathrm rates} + \chi^2_{\mathrm spectrum} 
\label{chitot}
\eeq
where
\begin{equation}
\chi^2_{\mathrm rates} =
\sum_{i,j} \left(F_i^{\mathrm th} -
F_i^{\mathrm exp}\right)
(\sigma_{ij}^{-2}) \left(F_j^{\mathrm th} - F_j^{\mathrm exp}\right).
\label{chir}
\end{equation}
Here $F_{i}^{\xi}= {T_i^{\xi}}/{T_{i}^{\mathrm SSM}}$, where $\xi$ designates
$th$ (for the theoretical prediction with oscillations) or $exp$
(for the experimental value) and $T_i$ stand for the quantities
total rates from different experiments.
We have used the weighted average of the three Ga experimental rates and 
thus we have 
4 data points for the total rates. 
The error
matrix $\sigma_{ij}$ contains the experimental errors,
theoretical errors and their correlations. 
The correlation matrix for the total rates is constructed as in 
\cite{fog95}. 
The logarithmic derivatives for the seismic model needed for the 
calculation of error correlation matrix are given in Table \ref{tab1}.
For the SSM we take the latest values from \cite{bp00}. 
The spectrum chi-square is defined as,
\begin{equation}
\chi^2_{\mathrm spectrum} =
\sum_{i,j} \left(X_{\mathrm sp} F_i^{\mathrm th} -
F_i^{\mathrm exp}\right)
(\sigma_{ij}^{-2}) \left(X_{\mathrm sp} F_j^{\mathrm th} - F_j^{\mathrm exp}\right).
\label{chisp}
\end{equation}
Here $F_{i}^{\xi}= {S_i^{\xi}}/{S_{i}^{\mathrm SSM}}$ where $\xi$ designates
$th$ (for the theoretical prediction with oscillations) or $exp$
(for the experimental value) and $S_i$ stands for the 
electron
energy spectrum for different energy bins. We have used 
1258 day SK data on the electron energy spectrum at day and night 
which includes the energy bin from 5.0-5.5 MeV also and 
we have 38 data points for the spectrum. 
$X_{\mathrm sp}$ is the normalisation factor for the spectrum 
which is floated as a free parameter in the global analysis in order to 
filter out the information on the total flux from the spectrum data. 
Thus  for the global analysis we have 41 degrees of freedom (DOF)
for the no oscillation scenario. 
For the error matrix in the spectrum data  we include the statistical error,
correlated and uncorrelated systematic errors and the error due to the 
calculation of the spectrum \cite{sksolarspec}. 
The no oscillation 
$\chi^2_{\mathrm min}$ for 41 DOF
for BP2000 SSM is 89.27 while for the seismic model 
it is 94.17.
  
Next, we perform the chi-square analysis assuming neutrino oscillations
to operate. For the oscillation analysis, there are two parameters
-- \dm and $\tan^2\theta$ and thus the number of DOF is 39. 
In Table \ref{tab5} we present the 
best-fit values of parameters, $\chi^2_{\mathrm min}$ and 
the goodness of fit (GOF) 
of the solutions for both BP2000 and seismic model 
for $\nu_e - \nu_{\mathrm active}$ transitions.
There are five allowed solution in both models --  Large Mixing Angle (LMA),
Vacuum Oscillation (VO), Low \dm - Quasivacuum Oscillation (LOW-QVO),
Just So$^2$ \cite{ragh} 
and Small Mixing Angle (SMA) -- in order of decreasing GOF. 
The GOF in these regions are more 
or less similar in both models.
The best-fit for both models comes in the 
LMA region.

\noindent                                                 
In order to understand the results of Table \ref{tab5},
we write  Eq. (\ref{probnight}) as
\beq
P_{ee} = P^{D} - \frac{1}{\epsilon}(2 P^{D} - 1) f_{\mathrm reg}
\label{pee}
\eeq
with $\epsilon = \cos2 \theta$, $f_{\mathrm reg} = |A_{2e}^{\oplus}|^2 - \sin^2\theta$
and $P^{D}$ defined above in Eq. (\ref{probday}).
In Fig. 1 we plot the regeneration factor $f_{\mathrm reg}$ and 
the actual Earth regeneration $R_{E} = P_{ee} - P^{D}$ 
vs. energy at the SK latitude for the best-fit values of
parameters in the SMA, LMA and LOW regions. 
Since the latitude of the other detectors are not very different
we do not expect $f_{\mathrm reg}$ and $R_{E}$ to be very different for these. 
Since $f_{\mathrm reg}$ is always positive, the possibility of regeneration
inside the Earth depends on $P_{J}$ 
For $P_{J} < \frac{1}{2}$ there is $\nue$ regeneration inside Earth
while for $P_{J} > \frac{1}{2}$ more $\nu_e$'s are flavor converted.

\begin{itemize}
\item For the SMA region $\epsilon \approx 1$ and from Fig. 1 we observe 
that $f_{\mathrm reg}$ is very small excepting for two peaks at
E $\approx$ 6 MeV and E $\approx$ 15 MeV corresponding to 
strong enhancement of the earth regeneration effect for the neutrinos 
passing through the core \cite{akh,petearth}. Hence 
\beq
P_{ee}^{\mathrm SMA} \approx P^D
\label{prsma}
\eeq
In this region
for low energy ($pp$) neutrinos, resonance is not encountered 
(resonance density $\gg$ maximum solar density) and hence $P_J\approx 0$  
and $\cos2\theta_{m} \approx 1$ giving $P_{ee}^{\mathrm SMA} \approx 1$.
For intermediate energy ($^{7}{Be}$) neutrinos
$\cos2\theta_{m} \approx -1$ (resonance density $\ll$ production density)
and $P_{ee}^{\mathrm SMA} \approx P_{J} \approx 0$ for these energies.
For high energy ($^8B$) neutrinos also, $\cos2\theta_{m} \approx -1$ and
$P_{ee}^{\mathrm SMA} \approx P_{J}$, with $P_J$ rising with energy. 
This energy dependence of the SMA survival 
probability gives a GOF of 
9.22\% for BP2000 and 9.21\% for the seismic model for a 
simultaneous description of  
all the four observed rates given 
in Table \ref{tab4} and the SK spectrum.

\item
For the $\dm$ of the LMA solution in Table \ref{tab5}, the motion of the 
neutrino in the solar matter is adiabatic for almost all neutrino 
energies and $P_J \approx 0$. 
For low energy neutrinos the matter effects are weak both 
inside the Sun and in Earth giving $f_{\mathrm reg} \approx 0$ and 
$\cos 2\theta_{m} \approx \epsilon$ so that for Ga energies 
\cite{concha} 
\beq
P_{ee}^{\mathrm LMA} \approx \frac{1}{2}(1 + \epsilon^2)
\label{prgalma}
\eeq
At SK and SNO energies matter effects result in $\cos2\theta_{m}\approx -1$ 
while $f_{\mathrm reg}$ is small 
but non-zero
($\approx$ 0.03 at 10 MeV as seen from fig. 1) 
giving
\begin{eqnarray}
P_{ee}^{\mathrm LMA} &\approx& \frac{1}{2}(1 - \epsilon) + f_{\mathrm reg}
\nonumber\\
&=& \sin^2\theta + f_{\mathrm reg}
\label{prsklma}
\end{eqnarray}
With the values of $\epsilon$ from Table \ref{tab5} 
and $f_{\mathrm reg}$ given in fig. 1
eqs. (\ref{prgalma}) and (\ref{prsklma}) 
approximately reproduce the rates of Table \ref{tab4}. 
Since the probability (\ref{prsklma})
is approximately energy independent it can account for the flat recoil 
electron energy spectrum. 
Since the seismic model predicts a higher value for 
observed to predicted ratio for the SK, SNO and Cl rates, 
the best-fit value of 
$\sin^2\theta$ or $\tan^2\theta$ obtained for seismic model are larger 
(cf. Eq. (\ref{prsklma})). 

\item
In the LOW region $\cos2\theta_m \approx -1$ for all neutrino energies 
and  $P_J\approx 0$ (except for very high energy neutrinos) and thus
\beq
P_{ee}^{\mathrm LOW} = \frac{1}{2}(1 - \epsilon) + f_{\mathrm reg}
\eeq
As is seen from Fig. 1 
$f_{\mathrm reg}$ is small  for high energy neutrinos and large for 
low energy neutrinos. 
For the best-fit case $\epsilon =0.2$ and $f_{\mathrm reg} \sim 0.2$ for 
Ga energies and $\sim 0.025$ for SK energies,
which can just about reconcile the Ga and SK 
rates. But it provides a very good description of 
the flat SK spectrum. 
The best-fit mixing angle is larger for the seismic case as
in LMA. 

\item  
At the best-fit value of \dm of Table \ref{tab5} for VO solutions 
the energy smearing over the bins washes out the energy variation due to the 
oscillations and 
the flat recoil electron spectrum observed at SK can be explained. 

\item 
For the \dm in the \js region one gets a very small survival
probability for the $^{7}{Be}$ neutrinos while for the \br
neutrinos the survival probability is close to 1.0 \cite{justso2}.
Therefore it cannot explain the total rates data but since it gives a 
flat probability for the \br  
neutrinos the spectrum shape can be accounted for and the global analysis gives a GOF of 8.1\% in BP2000 and 12.49\% 
in seismic.
Since the ratios of observed rates to predicted rates are higher for seismic
Just So$^2$ 
gives a lower contribution to $\chi^2_{\mathrm rates}$  and a better 
GOF in seismic.

\end{itemize}

In Fig. 2 we plot the 90\% ($\chi^2 \leq \chi^2_{\mathrm min}+4.61$), 95\%
($\chi^2 \leq \chi^2_{\mathrm min} +5.99$), 99\% ($\chi^2 \leq
\chi^2_{\mathrm min} +9.21$) and 99.73\% C.L.($\chi^2 \leq
\chi^2_{\mathrm min}+11.83)$ 
allowed areas in the $\dm-\tan^2\theta$ plane for both BP2000 
and the seismic model. 
We plot the allowed regions 
with respect to the {\it global minimum}.  
Both the models do not admit any allowed area in the SMA region.
For other regions the 
allowed areas in SSM and seismic model are roughly similar with 
the following differences observed in the seismic case:
\begin{itemize}
\item more area seems to be allowed in the LMA and LOW-QVO region
\item the LMA region extends to higher $\dm$
\item  higher values of mixing angles are allowed for LMA and
LOW-QVO regions, specifically, with LMA
extending into the dark zone ($\theta > \pi/4$).
\end{itemize}

For the Ga experiment, the 
seismic model predicts a higher $pp$ and lower $^{8}{B}$
flux as compared to BP2000. The net effect is that the increase of 
the flux ratio is smaller in Ga than 
in SNO, SK and Cl and the observed to predicted rate in 
Ga is closer to that in Cl, SK and SNO for the seismic model. Thus the 
energy dependence between the low energy $pp$ and high 
energy \br rates  
seen for BP2000 is reduced in the seismic model 
and the data can now be better accounted by an energy independent 
scenario. In fact, since the energy distortion in the observed rate 
is less in the seismic case, most parts of 
the bands in the parameter space with 
energy dependence of $<10\%$ are permitted\footnote{
Figure 1 of reference \cite{dp}.}. This accounts for the extended allowed 
areas appearing in Fig. 2 for the 
seismic model. These extended areas with weak energy dependence 
include both high \dm zones as well as high mixing angles\footnote{ 
The increase of the mixing angle with the decrease of the gap 
between the Ga and SK-SNO rates can also be inferred by comparing 
Eq. (\ref{prgalma}) and (\ref{prsklma}).}. In fact, 
the allowed area expands well into the dark zone for the LMA solution. 
For these values of the mixing angles the predicted 
flux ratio for the SK and SNO is more than that for Ga, against the 
energy trend of the data. However, for the seismic model 
these zones are still allowed at 99.73\% C.L. 
owing to the proximity of the Ga and SK-SNO rates.

In Fig. 3a and Fig. 3b we show the experimental rates with their $3\sigma$ 
errors, together with the $3\sigma$ range of predicted rates for the 
LMA (Fig. 3a) and LOW-QVO-VO (Fig. 3b) solutions, 
in the plane of any two experiments taken together. 
From the figures it is clear that the LMA (LOW-QVO) region can 
better account for the 
SSM(seismic model) predicted rates. 

In Table \ref{tab6} we give the $\chi^2_{\mathrm min}$ and the best-fit points for 
$\nue$ transition to sterile neutrino.  
Since the \br flux measured by the charged current reaction at SNO 
is significantly lower than the observed SK flux, all the sterile 
solutions appear to be disfavored with more than 90\% probability, except 
for the VO solution for BP2000 and both VO and Just So$^2$ for the 
seismic model. The VO solution produces better fits than the MSW solutions as 
it gives a lower contribution to the 
$\chi^2_{\mathrm spectrum}$ (cf.eq. (\ref{chitot}). 
As in the $\nu_e - \nu_{active}$ case, \js 
gives a better fit for seismic.

\section{Conclusions}
The measured solar neutrino fluxes have been consistently lower than
the theoretical predictions from SSM. 
We have constructed a seismic model of the sun consistent with the 
helioseismic data that 
predicts \br fluxes
lower than that predicted by BP2000, but the corresponding pp flux 
turns out to be slightly 
higher in the seismic model.
We examine how the use of the seismic model 
fluxes changes quality of the fits 
in the MSW and the vacuum oscillation region as 
compared to BP2000. 
For the statistical analysis of the data we use a $\chi^2$ minimization 
technique where we vary the 
normalization of the spectrum as a free parameter and 
thus avoid the over-counting of the SK observed flux and 
consider only the shape information from the SK spectral data.
We find that the inclusion of 
theoretical uncertainties and their correlations self consistently in both 
models result in fairly
similar GOF for the oscillation solutions in both models. 
However, we note that the use of seismic fluxes does modify the 
allowed areas in the parameter space. 
The 
increased proximity of the Ga and SK rates reduces the energy 
distortion of the observed fluxes and allows the regions of 
parameter space with weak energy dependence. 

\begin{acknowledgments}
This work utilizes data from the Solar Oscillations
Investigation/ Michelson Doppler Imager (SOI/MDI) on the Solar and
Heliospheric Observatory (SOHO). The MDI project is supported by NASA
grant NAG5-8878 to Stanford University. SOHO is a project of international
cooperation between ESA and NASA.
SMC wishes to express his thanks to the Leverhulme Trust for the
award of a Visiting Professorship at Queen Mary College, London. 
We thank Prof. S.T. Petcov for his comments.
\end{acknowledgments}

\begin{table}[htbp]
  \begin{center}
    \caption{\label{tab1}Neutrino fluxes in seismic model}
\vspace{1em}
    \begin{tabular}[h]{lcccccccc}
      \hline \\[-5pt]
Neutrino&Flux, $F_\nu$&\multispan7{\hfil Logarithmic derivatives
${\partial \ln F_\nu \over \partial \ln X}$ \hfil}\\
&(cm$^{-2}$ s$^{-1}$)&$S_{3,3}$&$S_{3,4}$&$S_{1,14}$&$S_{1,7}$&
$L_\odot$&$Z$&$\kappa$\\[+5pt]
      \hline \\[-5pt]
pp&$(6.12\pm0.01)\times10^{10}\!\!$&$\ph0.03$&$-0.07$&$-0.02$&$0.00$&$0.96$&$-0.
06$&$-0.13$\\
pep&$(1.45\pm0.01)\times10^8$&$\ph0.03$&$-0.07$&$-0.02$&$0.00$&$0.76$&$-0.17$&$-
0.31$\\

hep&$(2.12\pm0.01)\times10^3$&$-0.41$&$-0.07$&$-0.01$&$0.00$&$0.09$&$-0.21$&$-0.
41$\\
$^7$Be&$(4.54\pm0.12)\times10^9$&$-0.44$&$\ph0.95$&$\ph0.00$&$0.00$&$2.06$&$\ph0
.68$&$\ph1.45$\\
$^8$B&$(4.16\pm0.26)\times10^6$&$-0.49$&$\ph1.04$&$\ph0.01$&$1.00$&$3.85$&$\ph1.
65$&$\ph3.40$\\
$^{13}$N&$(5.02\pm0.25)\times10^8$&$-0.08$&$\ph0.14$&$\ph0.85$&$0.00$&$2.39$&
$\ph 1.24$&$\ph2.52$\\
$^{15}$O&$(4.14\pm0.26)\times10^8$&$-0.09$&$\ph0.17$&$\ph1.00$&$0.00$&$2.93$&
$\ph1.52$&$\ph3.07$\\
$^{17}$F&$(4.77\pm0.32)\times10^6$&$-0.10$&$\ph0.18$&$\ph0.01$&$0.00$&
$3.07$&$\ph1.59$&$\ph3.21$\\
\noalign{\medskip}
Rel. error&&0.060&0.094&0.143&0.11&.004&0.033&0.02\cr
      \hline \\
      \end{tabular}
  \end{center}
\end{table}

\begin{table}
  \begin{center}
    \caption{\label{tab2}Predicted neutrino fluxes in various solar
neutrino experiments}
\vspace{1em}
    \begin{tabular}[h]{lllll}
      \hline \\[-5pt]
Experiment&Homestake&SK&$\!\!\!\!$Gallex, SAGE, GNO$\!\!\!\!$ &SNO CC\\
&(${^{37}}$Cl)&(H${_2}$O)&(${^{71}}$Ga)&(D${_2}$O)\\
\noalign{\smallskip}
&(SNU)&$\!\!\!\!({10^6}$cm${^{-2}}$s${^{-1}})\!\!$&(SNU)&
(${10^6}$cm${^{-2}}$s${^{-1}}$)\\[+5pt]
      \hline \\[-5pt]
Measured Flux&${2.56\pm0.23}$&${2.32\pm0.08}$&${74.7\pm5.0}$
& $1.75 \pm 0.14$\\[+6pt]
SSM (BP00)&${7.6^{+1.3}_{-1.1}}$&${5.05^{+1.01}_{-0.81}}$
&${128.0^{+9}_{-7}}$ & ${5.05^{+1.01}_{-0.81}}$\\
SSM (BP95)&${9.3^{+1.2}_{-1.4}}$&${6.62^{+0.93}_{-1.13}}$&${137^{+8}_{-7}}$
&${6.62^{+0.93}_{-1.13}}$\\
SSM (BP98)&${7.7^{+1.2}_{-1.0}}$&${5.15^{+0.51}_{-0.72}}$&${129^{+8}_{-6}}$
&${5.15^{+0.51}_{-0.72}}$\\
SSM (Bru98)&${7.2}$&${4.8}$&${127}$&${4.8}$\\
SSM (Bru99)&${6.7}$&${4.7}$&${125}$&${4.7}$\\
Seismic model &${6.46\pm0.99}$&${4.16\pm0.76}$&${124.9\pm6.5}$
&${4.16\pm0.76}$\\
      \hline 
      \end{tabular}
  \end{center}
\end{table}

\begin{table}[htbp]
      \caption{\label{tab3}}
    The seismic model predictions for the solar neutrinos
    fluxes and neutrino capture rates in the Cl and Ga detectors.
    The expected $^{8}$B flux in SK and SNO is as given in Table \ref{tab2}.

    \begin{center}
        \begin{tabular}{cccccc} \hline
          &\multicolumn{2}{c}{seismic}&\multicolumn{2}{c}{BP2000}\\
         \cline{2-5}
         {\rule[-3mm]{0mm}{8mm}
         source}  & $^{37}$Cl & $^{71}$Ga &$^{37}$Cl & $^{71}$Ga &\\
    [1ex]      & (SNU) & (SNU)& (SNU) & (SNU)\\
    [1ex]\hline
         {\rule[-3mm]{0mm}{8mm}
         $pp$}  &  0.00 & 71.58 &  0.00 & 69.7\\[1ex]
         $pep$ &   0.23 & 2.95 & 0.22 & 2.8\\[1ex]
         $hep$ &   0.04 & 0.07 & 0.04 & 0.1 \\[1ex]
         $^{7}$Be & 1.09 & 32.57 & 1.15 & 34.2 \\[1ex]
         $^8$B &  4.74 & 9.98 & 5.76 & 12.1 \\[1ex]
         $^{13}$N &  0.09 & 3.03 & 0.09 & 3.4 \\[1ex]
         $^{15}$O &0.27 & 4.72 & 0.33 & 5.5 \\[1ex]
         $^{17}$F &  0.00 & 0.05 & 0.00 & 0.1 \\ [1ex]\hline
                 & & \\
         \raisebox{1.5ex}[0pt] {Total} &  
         \raisebox{1.5ex}[0pt] {6.46}&
         \raisebox{1.5ex}[0pt] { 124.94} 
         &\raisebox{1.5ex}[0pt] {$7.6$}&
         \raisebox{1.5ex}[0pt] { $128.0$} \\
         \hline

         \end{tabular}

      \end{center}
\end{table} 

\begin{table}[htbp]
{\footnotesize{
\caption{\label{tab4} The observed rates relative to the theoretical 
predictions
for Ga, Cl, SK and SNO experiments along with their
compositions(comp.) for both seismic model and SSM. 
The Ga rate corresponds to the combined
SAGE and GALLEX+GNO data.}
\let\tabbodyfont\footnotesize
    \begin{center}
        \begin{tabular}{cccccc}\hline
         {\rule[-3mm]{0mm}{10mm}
          Model} & Experiment & Ga & Cl & SK & SNO (CC) \\ \hline
          & Rate & $0.584 \pm 0.039$ & $0.337\pm0.030$& $0.459\pm 0.017$ 
            & $0.347 \pm 0.027 $\\
          BP2000&    &                  &    & \\
         &Comp.  & $pp$ (55\%), \ber (25\%), \br (10\%) &
         \ber (15\%), \br (75\%) & \br (100\%) & \br (100\%)\\ \hline
          & Rate & $0.598 \pm 0.040$ & $0.396\pm0.035$ & $0.557\pm 0.021$ 
           & $0.421\pm0.033$\\
          SEISMIC &    &     &    &  & \\
         &Comp.  & $pp$ (57\%), \ber (26\%), \br (8\%) &
         \ber (17\%), \br (73\%) & \br (100\%) & \br (100\%) \\ \hline
        \end{tabular}
      \end{center}
}} 
\end{table}

\begin{table}
\caption{\label{tab5}
{The best-fit values of the parameters,
$\chi^2_{\mathrm min}$, and the goodness of fit from the
global analysis of
rates and the 1258 day SK day-night spectrum data for MSW analysis involving
two active neutrino flavors.}}
 \begin{center}
  \begin{tabular}{cccccc}
   \hline
   &Nature of & $\Delta m^2$ &
   $\tan^2\theta$&$\chi^2_{\mathrm min}$& Goodness\\
   &Solution & in eV$^2$&  & & of fit\\
   \hline
   &SMA & $5.28 \times 10^{-6}$&$3.75 \times 10^{-4}$ &
   51.14 & 9.22\%  \\
 &LMA & $4.70 \times 10^{-5}$ & 0.38 &
   33.42 &  72.18\% \\
    BP2000
   &LOW-QVO & $ 1.76 \times 10^{-7}$ & 0.67 & 39.00 & 46.99\%
   \\
    &VO& $4.64\times 10^{-10}$& 0.57 & 38.28 & 50.25\%\\
    &Just So$^2$&$5.37\times 10^{-12}$& 0.77&51.90&8.10\%\\
    \hline
   &SMA & $4.66 \times 10^{-6}$ & $5.10 \times 10^{-4}$ &
   51.15 & 9.21\%  \\
   &LMA & $5.11 \times 10^{-5}$ & 0.44 &
   35.62 & 62.48\%  \\
   SEISMIC
   &LOW-QVO &$1.76 \times 10^{-7}$&$0.71$ &
   38.22& 50.53\%  \\
    & VO & $4.65 \times 10^{-10}$& 0.47 & 36.23 & 59.69\% \\
    &Just So$^2$&$5.37\times 10^{-12}$& 1.00&49.30&12.49\%\\
   \hline
  \end{tabular}
  \end{center}
\end{table}

\begin{table}
\caption{\label{tab6}
{The The best-fit values of the parameters,
$\chi^2_{\mathrm min}$, and the goodness of fit from the
global analysis of
rates and the 1258 day SK day-night spectrum data for two-generation
$\nu_e-\nu_{\mathrm sterile}$ MSW analysis.}}
 \begin{center}
  \begin{tabular}{cccccc}
   \hline
   &Nature of & $\Delta m^2$ &
   $\tan^2\theta$&$\chi^2_{\mathrm min}$& Goodness\\
   &Solution & in eV$^2$&  & & of fit\\
   \hline
   &SMA & $5.59 \times 10^{-6}$&$2.83 \times 10^{-4}$ &
   54.21 & 5.35\%  \\
   &LMA & $6.13 \times 10^{-5}$ & 0.50 &
   52.93 &  6.75\% \\
   BP2000
   &LOW-QVO & $ 2.93 \times 10^{-8}$ & 1.00 & 53.18 & 6.46\% \\
    &VO& $4.67\times 10^{-10}$& 0.37 & 46.28 & 19.70\%\\
    &Just So$^2$& $5.37\times 10^{-12}$& 0.77 & 52.09 & 7.83\%\\
    \hline
   &SMA & $3.81 \times 10^{-6}$ & $3.67 \times 10^{-4}$ &
 58.18 & 2.47\%  \\
   &LMA & $6.04 \times 10^{-5}$ & 0.62 &
   54.97 & 4.64\%  \\
     SEISMIC
   &LOW-QVO &$3.20 \times 10^{-8}$&$1.00$ &
   53.26& 6.36\%  \\
    & VO & $4.68 \times 10^{-10}$& 0.37 & 44.82&24.09\% \\
    &Just So$^2$& $5.37\times 10^{-12}$& 0.98 & 49.51 & 12.07\%\\
   \hline
 \end{tabular}
\end{center}
\end{table}

\newpage
\topmargin -1in
\centerline{\psfig{figure=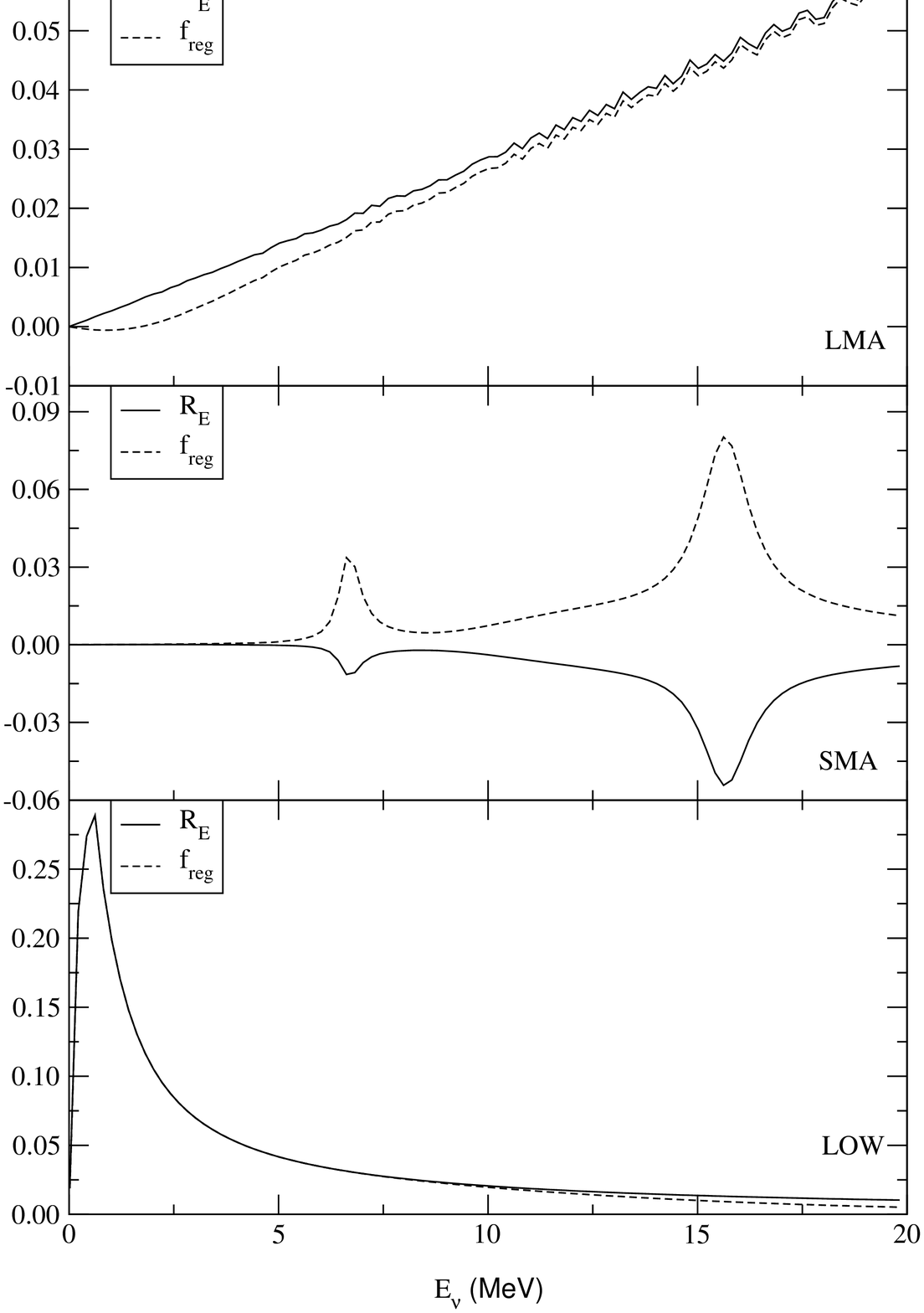,width=15.5cm,height=18.5cm}}
\vskip 0.25in
\parbox{6in}{
{\bf Fig. 1}: The regeneration factor $f_{\mathrm reg} = 
(|A_{2e}|^2 - \sin^2 \theta$)
and the net Earth regeneration 
$R_{E} (=P_{ee} - P^{D})$ as a function of energy for 
typical values of the parameters in the SMA, LMA and LOW-QVO regions.}

\newpage
\topmargin -1in
\centerline{\psfig{figure=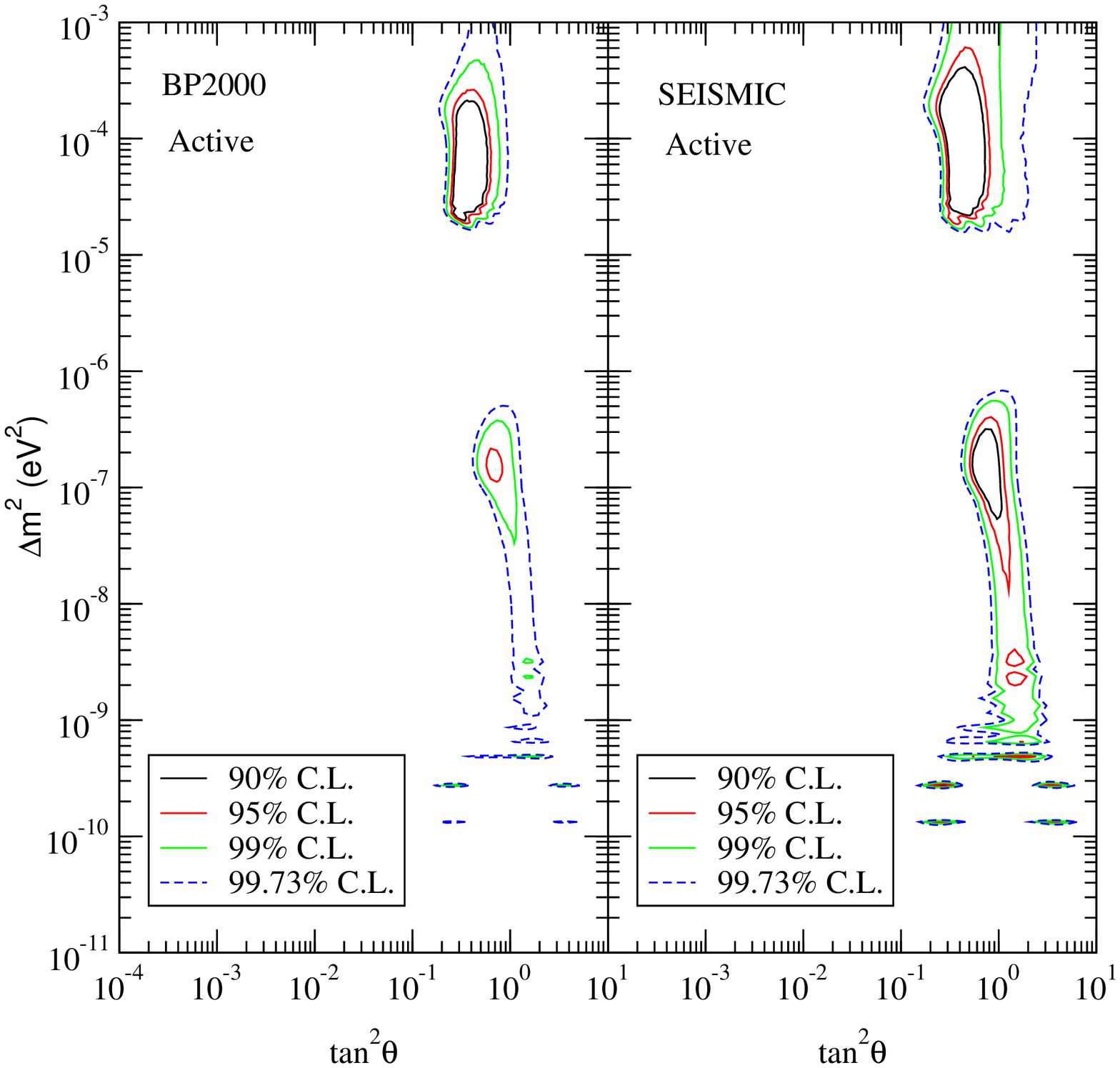,width=16.5cm,height=20.5cm}}
\vskip -1in
\parbox{6in}{
{\bf Fig. 2}: 
The 90\%, 95\%, 99\% and  99.73\% C.L. allowed areas from the
global analysis of the total rates from Cl, 
Ga, SK and SNO detectors
and the 1258 day SK recoil electron spectrum at day and night,
assuming MSW conversions to active neutrinos, using theoretical predictions 
from BP2000 and seismic model.}

\newpage
\topmargin -1in
\centerline{\psfig{figure=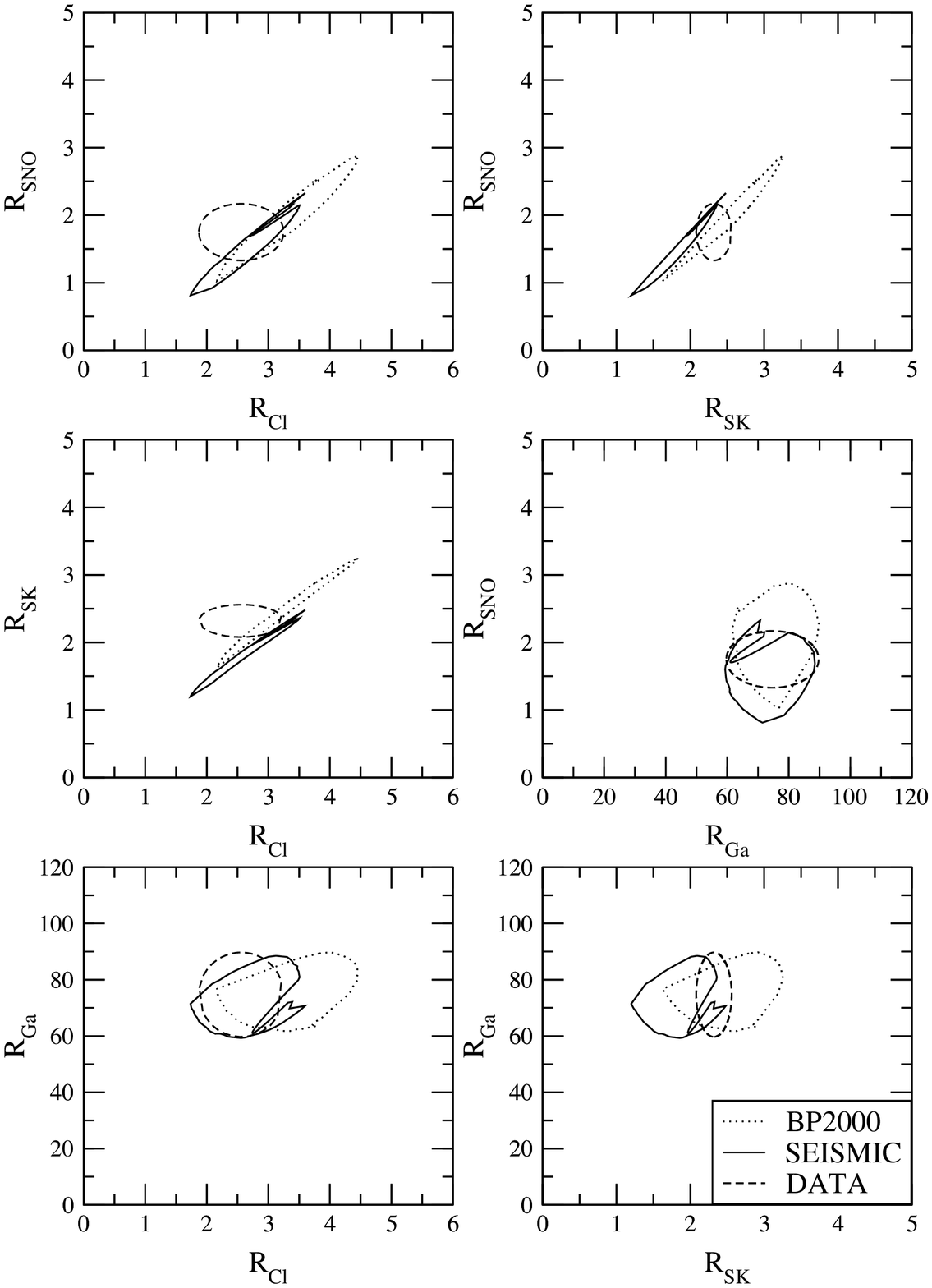,width=14.5cm,height=18.5cm}}
\parbox{6in}{
{\bf Fig. 3a}: The experimental rates (in SNU for the Cl and Ga experiments 
and in units of $10^6$ cm$^{-2}$s$^{-1}$ for SK and SNO) 
with $3\sigma$ errors 
(shown by ellipses) and the 99.73\% C.L. range of predicted rates for 
LMA solution, in the plane of any two of the experiments
for BP2000 (dotted line) and 
for seismic model (solid line).}

\newpage
\topmargin -1in
\centerline{\psfig{figure=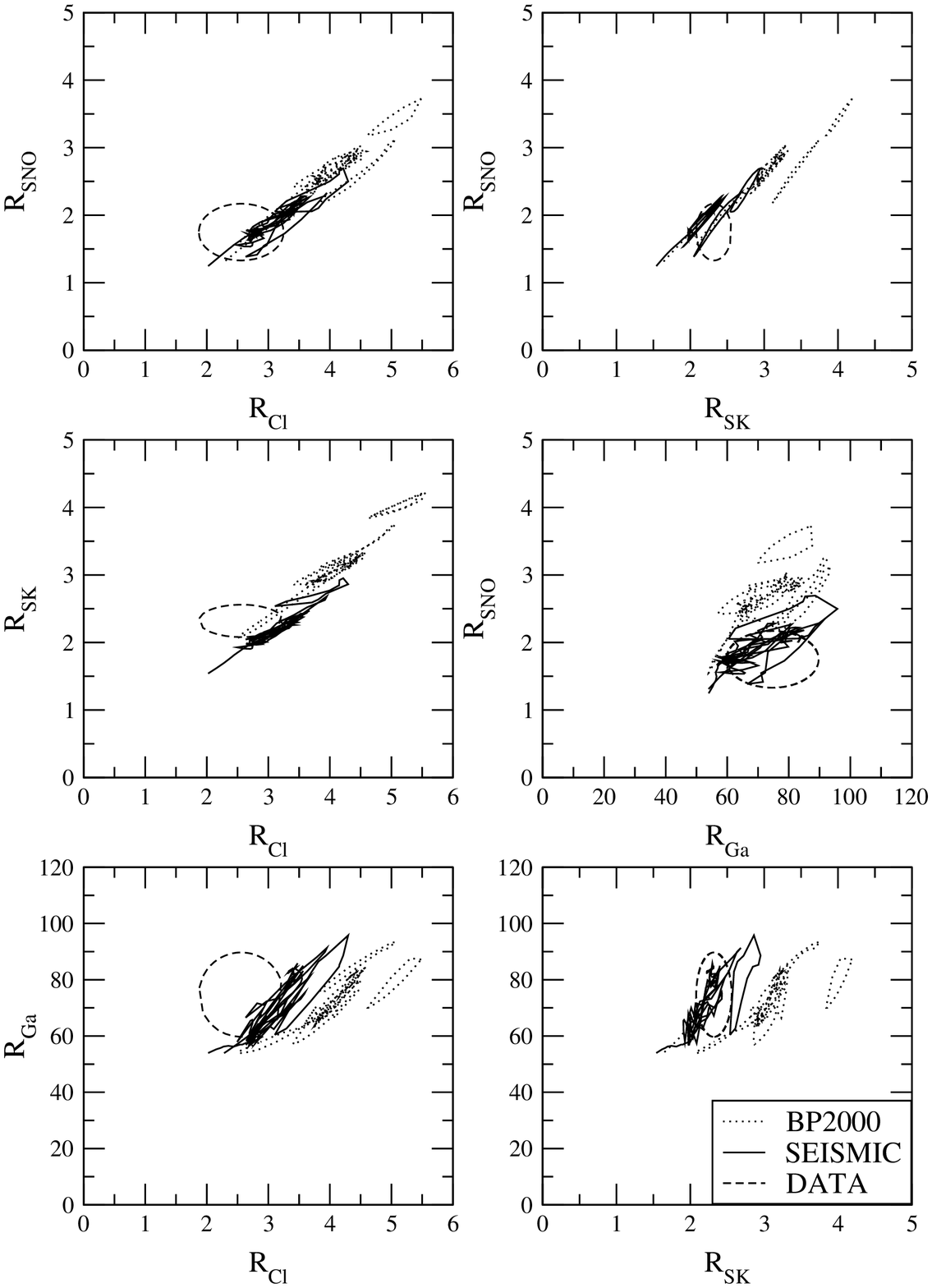,width=14.5cm,height=18.5cm}}
\parbox{6in}{
{\bf Fig. 3b}: 
same as in fig. 3a but for LOW-QVO-VO solution. 
For this solution there are multiple contours
(cf. Fig.~2), resulting in a complicated pattern.}

\end{document}